\documentclass[12pt]{article}
\begin{document}
\begin{center}
{\large \bf Multi-Hamiltonian structure of \\ Plebanski's second heavenly equation}\\[15mm]
{\large F. Neyzi$^*$, Y. Nutku$^\dagger$} and {\large M. B. Sheftel$^*$}\\[2mm]
$^*$ Dept. of Physics, Bo\u{g}azi\c{c}i University,
Bebek, Istanbul  34342 Turkey \\
$^\dagger$ Feza G\"ursey Institute P.O.Box 6 \c{C}engelk\"oy, Istanbul 81220 Turkey\\[10mm]
\end{center}

\vspace{5mm}

\noindent Email: neyzif@boun.edu.tr, nutku@gursey.gov.tr,\\
mikhail.sheftel@boun.edu.tr, sheftel@gursey.gov.tr

 \vspace{1cm}

\noindent  {\bf Abstract}

\noindent We show that Plebanski's second heavenly equation, when
written as a first-order nonlinear evolutionary system, admits
multi-Hamiltonian structure. Therefore by Magri's theorem it is a
completely integrable system. Thus it is an example of a
completely integrable system in four dimensions. \vspace{1cm}



\section{Introduction}

The Einstein field equations that govern self-dual gravitational
fields reduce to a single scalar-valued equation. This is either
the complex Monge-Amp\`{e}re equation that Calabi \cite{calabi}
had shown to govern Ricci-flat K\"ahler metrics and which
Plebanski
 \cite{pleb} has called the first heavenly equation, or
\begin{equation}
u_{tt} u_{xx} - u_{tx}^{\;\;\;2} + u_{xz} + u_{ty} = 0 \, ,
\label{heaven2}
\end{equation}
which is Plebanski's second heavenly equation. In this paper we
shall consider the Hamiltonian structure of the second heavenly
equation. We shall show that it can be formulated as a Hamiltonian
system in three or more inequivalent ways. Therefore, by the
theorem of Magri \cite{magri} it is a completely integrable
system. The only known example of a completely integrable system
in four dimensions was anti-self-dual Yang-Mills fields. Here we
show that the second heavenly equation which governs
anti-self-dual gravitational fields is another example of a
completely integrable system in four dimensions.

The real Monge-Amp\`{e}re equation admits bi-Hamiltonian structure
\cite{nrma2} and in the case of the complex Monge-Amp\`{e}re
equation we have two symplectic structures \cite{yn} {\it except}
in the crucial case of two complex dimensions. It is natural to
expect that the second heavenly equation (\ref{heaven2}) may admit
bi-Hamiltonian structure. This expectation is further supported by
the general feeling that self-dual gravitational fields should be
integrable systems.

Earlier Dunajski and Mason \cite{dm2} mentioned briefly that
Plebanski's second heavenly equation may admit bi-Hamiltonian
structure. For this purpose they rewrote (\ref{heaven2}) as a
nonlocal first order equation. They used a scalar recursion
operator appropriate to the original one-component second heavenly
equation. They did not explicitly construct Hamiltonian operators
determining the structure of Poisson brackets, nor the family of
commuting Hamiltonians. In their later paper \cite{dm} Dunajski
and Mason suggested another recursion operator for the
one-component equation (\ref{heaven2}) determined by two
different, though compatible, recursion relations but they did not
develop the Hamiltonian formalism there.

We shall consider a different, a two-component representation of
the second heavenly equation as a first order system of two {\it
local} equations and show that Plebanski's second heavenly
equation indeed admits Magri, {\it i.e.} multi-Hamiltonian,
structure. Earlier we had constructed \cite{mns} the scalar
recursion operator for the second heavenly equation and now we
shall cast it into $2\times 2$ matrix form which naturally joins
the two recursion relations into one matrix relation.

In section \ref{sec-1storder} we introduce a first-order
two-component form of the second heavenly equation
(\ref{heaven2}). In section \ref{sec-hamilton} we present the
first Hamiltonian structure of this system of equations. We start
with a degenerate Lagrangian and construct its Dirac bracket
\cite{dirac} to find the Hamiltonian operator. In section
\ref{sec-symplect} we invert the Hamiltonian operator and obtain
the corresponding symplectic 2-form. In section \ref{sec-recurs}
we construct explicitly a matrix integral-differential recursion
operator in the two-component form which incorporates naturally
both recursion relations. This operator and the operator
determining symmetries form a Lax pair for the two-component
system. In section \ref{sec-multihamilton} we give explicitly the
second and the third Hamiltonian structures which show the way for
obtaining multi-Hamiltonian representation of our two-component
system. In section \ref{sec-symmetry} some simple symmetries of
our equations are shown to be generated by certain integrals of
motion via previously constructed Hamiltonian operators. We find
the complete Lie algebra of point symmetries of the two-component
system and for all variational symmetries we construct the
corresponding integrals of motion. In section \ref{sec-higher} we
give examples of higher flows obtained with the aid of the
Hermitian conjugate of the recursion operator. Those flows are
nonlocal symmetries generated by local integrals.

\section{First order form of the second heavenly equation}
\label{sec-1storder}

   The second heavenly equation is a second order partial
differential equation. In order to discuss its Hamiltonian
structure we shall single out an independent variable, $t$, in
(\ref{heaven2}) to play the role of ``time" and express the second
heavenly equation as a pair of first order nonlinear evolution
equations. Thus we introduce an auxiliary variable $q$ whereby
(\ref{heaven2}) assumes the form
\begin{equation}
u_{t}  =    q\; ,   \;\;\;\;\;\;\;
q_{t}  =  \frac{\textstyle{1}}{\textstyle{u_{xx}}} \left(
        q_{x}^{\,2}  - q_{y} - u_{xz} \right) \equiv Q
\label{uq}
\end{equation}
of a first order system. For the sake of brevity we shall
henceforth refer to (\ref{uq}) as the second heavenly system. It
is worth noting that this split of (\ref{heaven2}) into the system
(\ref{uq}) is not unique, here we are using the most
straight-forward choice. The choice of the independent variable
$t$ as ``time" is arbitrary and in no way represents any sort of a
physical time variable. Also we shall henceforth assume $u_{xx}
\ne 0$. This is not an essential restriction but rather a
statement of non-triviality. Now the vector field
\begin{equation}
{\bf X} = q   \frac{\partial}{\partial u} + Q
 \frac{\partial}{\partial q}
\label{uqvector}
\end{equation}
defines the flow. In the discussion of the Hamiltonian structure
of this system we shall use matrix notation with $$ u^i \;\;\;
i=1,2 \qquad u^1=u, \;\; u^2 = q $$ running over the dependent
variables.

The equations of motion (\ref{uq}) are to be cast into the form of
Hamilton's equations in two different ways according to the
recursion relation of Magri
\begin{equation}
u^i_t = {\bf X}( u^i) = J^{ik}_{0} \; \delta_k H_{1} = J_1^{ik} \;
\delta_k H_{0} \label{hameqpq}
\end{equation}
where $\delta_k$ denotes variational derivative of the Hamiltonian
functional with respect to $u^k$.

\section{First Hamiltonian structure}
\label{sec-hamilton}

There is a systematic way to derive the first Hamiltonian
structure of (\ref{uq}). It was used to obtain Hamiltonian
structures for the real Monge-Amp\`ere equation \cite{nrma2}. We
shall now apply it to the second heavenly system.

We start with an action principle for (\ref{uq}) given by the
Lagrangian density
\begin{equation}
{\cal L} = q \, u_t\, u_{xx} + \frac{1}{2}\, u_t \, u_y -
\frac{1}{2}\, q^2 \, u_{xx} + \frac{1}{2} \, u_{x} \, u_{z}
\label{lag}
\end{equation}
which is degenerate. We need to apply Dirac's theory of
constraints \cite{dirac} in order to arrive at its Hamiltonian
formulation. We define canonical momenta
$$ \pi_i = \frac{\partial L}{\partial u^i_t}$$
which satisfy canonical Poisson brackets
$$[ \pi_i(\xi), u^k(\eta) ] = \delta^k_i \, \delta(\xi-\eta) $$
where $\xi, \eta$ are generic names for independent variables each
one of which stands for the collection of our original independent
variables $x,y,z$. In other words, $\delta(\xi-\eta) \equiv
\delta(x-x') \, \delta(y-y') \,  \delta(z-z')$ for $\xi =
\{x,y,z\}, \eta = \{x',y',z'\}$. We note that the momenta cannot
be inverted for the velocities because the Lagrangian (\ref{lag})
is degenerate. Therefore following Dirac \cite{dirac} we impose
them as constraints
\begin{eqnarray}
\phi_u & = & \pi_u - (q \, u_{xx} + \frac{1}{2}\, u_y)
\label{constraints}\\
\phi_q & = & \pi_q \nonumber
\end{eqnarray}
and calculate the Poisson bracket of the constraints
\begin{equation}
K_{ik} = \left[ \phi_i(x,y,z) , \phi_k(x',y',z') \right]
\label{kik}
\end{equation}
organizing them into the form of a matrix. We find
\begin{eqnarray}
\left[ \phi_u(x,y,z) , \phi_u(x',y',z') \right] & = & - \, q(x')
\,
\delta_{x' x'}(x-x') \, \delta(y-y') \, \delta(z-z') \nonumber\\
& & + \, q(x) \,
\delta_{x x}(x-x') \, \delta(y-y') \, \delta(z-z') \nonumber\\
&  & -
\delta(x-x') \, \delta_y(y-y') \, \delta(z-z') \label{pbconstraints} \\
\left[ \phi_u(x,y,z) , \phi_q(x',y',z') \right] & = & - u_{xx}\,
\delta(x-x') \, \delta(y-y') \, \delta(z-z')   \nonumber \\
\left[ \phi_q(x,y,z) , \phi_q(x',y',z') \right] & = & 0\nonumber
\end{eqnarray}
and see that they do not vanish modulo the constraints. Thus the
constraints are second class in the terminology of Dirac. We need
to invert the matrix of Poisson brackets of the constraints in
order to arrive at the Dirac bracket. That is why we have not
simplified (\ref{pbconstraints}) using properties of the Dirac
delta-function \cite{galvao}. We shall do so in sequel,
(\ref{kmu}), but it is better to keep them in raw form for
purposes of inversion because the definition of the inverse
\begin{equation}
\int K_{ik} (\xi, \sigma) J^{kj} (\sigma, \eta) \, d \sigma =
\delta_{i}^{j} \, \delta ( \xi - \eta ) \label{mik}
\end{equation}
results in a set of differential equations for the entries of
$J^{kj}$.

Given any two smooth functions of the canonical variables ${\cal
A}, {\cal B}$ the Dirac bracket is defined by \cite{dirac}
\begin{equation} \begin{array}{ll}
[ {\cal A}(\xi), {\cal B}(\eta) ]_{D} = & [ {\cal A}(\xi), {\cal B}(\eta) ] \\[4mm]
      & - \, \displaystyle{\int} [ {\cal A}(\xi), \phi_i(\theta) ]
J^{ik}(\theta, \sigma)  [ \phi_k(\sigma) , {\cal B}(\eta) ] \, d
\theta \, d \sigma
\end{array}
\label{defdirac}
\end{equation}
where $J^{ik}$ is the inverse of the matrix of Poisson brackets of
the constraints defined by (\ref{mik}).

    The first Hamiltonian operator for second heavenly system is
given by the Dirac bracket
\begin{equation}
J^{ik}_{0} = \left(            \begin{array}{cc}
 0 & \frac{ {\textstyle  1} }{ {\textstyle u_{xx} } } \\[4mm]
- \frac{ {\textstyle 1}}{  {\textstyle u_{xx} } }
 & \begin{array}{c}  \frac{ {\textstyle q_x }}{{\textstyle u_{xx}^{\;\;2} }} D_x
+ D_x  \frac{ {\textstyle q_x }}{{\textstyle u_{xx}^{\;\;2} }} \\-
\frac{ {\textstyle  1} }{ {\textstyle u_{xx} } } D_{y}
 \frac{ {\textstyle  1} }{ {\textstyle u_{xx} } } \end{array}
\end{array}  \right)               \label{j0}
\end{equation}
and it can be directly verified that
\begin{equation}
{\cal H}_1 = \frac{1}{2}\, q^2 \, u_{xx} -  \frac{1}{2} u_{x}
u_{z} \label{h1}
\end{equation}
is the conserved Hamiltonian density for the flow (\ref{uq}). In
other words, the total $t$-derivative of ${\cal H}_1$ along the
flow (\ref{uq}) is a three-dimensional total divergence in
``space" variables
\[\frac{\partial{\cal H}_1}{\partial t} = \frac{\partial P_1}{\partial x}
+ \frac{\partial Q_1}{\partial y} + \frac{\partial R_1}{\partial
z}\] and hence the total $t$-derivative of the functional
$H_1=\int_{-\infty}^\infty {\cal H}_1 dx dy dz$ vanishes by the
divergence theorem, provided the components of the current $P_1,
Q_1, R_1$ approach zero fast enough as $x, y, z$ go to
$\pm\infty$.

The Hamiltonian functional obtained from the density (\ref{h1}) is
the one that yields the equations of motion (\ref{uq}) by the
action of the operator (\ref{j0}).

The proof  of the Jacobi identities for the Hamiltonian operator
(\ref{j0}) is straight-forward but rather lengthy. A shorter proof
is obtained by inverting (\ref{j0}) to arrive at the symplectic
structure of the second heavenly system.

\section{Symplectic structure}
\label{sec-symplect}

The statement of the symplectic structure of the equations of
motion (\ref{uq}) consists of
\begin{equation}
 i_X \omega  =  d H   \label{symhameq}
\end{equation}
which is obtained by the contraction of the closed symplectic
2-form $\omega$ with the vector field {\bf X} (\ref{uqvector})
defining the flow. The symplectic 2-form is obtained by
integrating the density
\begin{equation}
\omega = \frac{1}{2} \, d u^i \wedge K_{ij} \, d u^j
\label{defomega}
\end{equation}
where $K$ is the inverse of $J_0$ given by (\ref{j0}). But we
already know it from the Poisson brackets of the constraints
(\ref{pbconstraints}). Thus the inverse of the Hamiltonian
operator (\ref{j0}) is given by
\begin{equation}
 K =    \left(            \begin{array}{ccc}
 q_{x} D_{x} + D_{x} q_{x} - D_{y}  & - u_{xx} \\
  u_{xx} & 0
\end{array}   \right)
\label{kmu}
\end{equation}
which is a local operator. Then we find the symplectic 2-form from
(\ref{defomega})
\begin{equation}
\omega =q_x \, d u \wedge d u_{x} - u_{xx} \,
 d u \wedge d q - \frac{1}{2} d u \wedge d u_{y}
\label{omu}
\end{equation}
which, up to a divergence, (\ref{omu}) can be directly verified to
be a closed 2-form.  The closure of the symplectic 2-form
(\ref{omu}) is equivalent to the satisfaction of the Jacobi
identities for the Hamiltonian operator (\ref{j0}).

\section{Recursion operator}
\label{sec-recurs}

Recently a recursion operator for second heavenly equation was
obtained \cite{dm}, \cite{mns}. We shall use it to construct new
Hamiltonian operators satisfying Magri's recursion relation
(\ref{hameqpq}). But before we can do so we need to express the
recursion operator in the two-component form appropriate to the
system (\ref{uq}).

We start with the equation determining the symmetries of the
second heavenly system. We introduce two components for symmetry
characteristics
\begin{equation} \begin{array}{c}
u_\tau = \varphi \\q_\tau = \psi \end{array} \qquad
 \Phi \equiv \left(
\begin{array}{c} \varphi \\ \psi \end{array} \right)
\label{fipsi}
\end{equation}
of the system (\ref{uq}). From the Frech\'et derivative of the
flow we find
\begin{equation}
 {\cal A} =  \left( \begin{array}{cc} D_t & - 1 \\ \frac{{\textstyle Q }}
 {{\textstyle u_{xx} }} D_x^2 +
\frac{{\textstyle 1 }}{{\textstyle u_{xx} }} D_x D_z
 & D_t - \frac{{\textstyle 2 q_x }}{{\textstyle u_{xx} }} D_x
 + \frac{{\textstyle 1 }}{{\textstyle u_{xx} }} D_y
 \end{array}
\right) \label{determiningop}
\end{equation}
and the equation determining the symmetries of the second heavenly
system is given by
\begin{equation}
 {\cal A}  \left( \Phi \right) = 0. \label{determiningeq}
\end{equation}
We note that the combination of the first and second determining
equations (\ref{determiningeq}) with the latter multiplied by an
overall factor of $u_{xx}$, coincides with the determining
equation for symmetries of the original second heavenly equation
(\ref{heaven2}). It was noted in \cite{mns} that this determining
equation has the divergence form
\begin{equation}
(u_{xx}\psi-q_x\varphi_x+\varphi_y)_t +
(q_t\varphi_x-q_x\psi+\varphi_z)_x = 0
\label{div}
\end{equation}
rewritten in two-component form. This implies the local existence
of the potential variable $\tilde\varphi$ such that
\begin{eqnarray}
\tilde\varphi_t &=& q_t\varphi_x-q_x\psi+\varphi_z \nonumber
\\ \tilde\varphi_x &=& -(u_{xx}\psi-q_x\varphi_x+\varphi_y)
\label{pot}
\end{eqnarray}
which, as is proven in \cite{mns}, satisfies the same determining
equation for symmetries of (\ref{heaven2}) and therefore is a
``partner symmetry" for $\varphi$ \cite{mnsbig}. In the
two-component form we define the second component of this new
symmetry similar to the definition of $\psi$ as $\tilde\psi =
\tilde\varphi_t$. Then the two-component vector
\[\tilde\Phi = \left(
\begin{array}{c} \tilde\varphi \\ \tilde\psi \end{array} \right)
\]
satisfies the determining equation for symmetries in the form
(\ref{determiningeq}) and hence is a symmetry characteristic of
the system (\ref{uq}) provided the vector (\ref{fipsi}) is also a
symmetry characteristic. Thus (\ref{pot}) become the recursion
relation for symmetries in the two-component form
\begin{equation}
\tilde\Phi
 = {\cal R} ( \Phi )
\label{recursrel}
\end{equation}
with the recursion operator given by
\begin{equation}
 {\cal R} =  \left( \begin{array}{cc} D_x^{\;-1} (q_x D_x - D_y)
   & -  D_x^{\;-1} u_{xx}  \\ Q D_x +  D_z
 &  - q_{x}
 \end{array}
\right)  \label{recursion}
\end{equation}
where $ D_{x}^{\;-1} $ is the inverse of $ D_{x}$. See
\cite{fokas} for the definition and properties of this operator,
in particular,
\begin{equation}
 D_x^{\;-1} f =  \frac{1}{2}\left( \int^x_{-\infty} -
\; \int_x^\infty\right) \,  f(\xi) \,  d \xi
\end{equation}
and the integrals are taken in the principal value sense. The
commutator of the recursion operator (\ref{recursion}) and the
operator determining symmetries (\ref{determiningop}) has the form
\begin{equation}
\left[ {\cal R} ,  {\cal A} \right] = \left(
\begin{array}{cc}
D_x^{-1}(q_t-Q)_{xx}-(q_t-Q)_x & D_x^{-1}(u_t-q)_{xx}
\\
\left\{\frac{\textstyle Q}{\textstyle
u_{xx}}(u_t-Q)_{xx}+(D_y-\frac{\textstyle 2q_x}{\textstyle u_{xx}}
)({\textstyle q_t-Q })_x\right\}D_x & (q_t-Q)_x
\end{array}
\right)
\label{comRA}
\end{equation}
and as a consequence, the operators ${\cal R}$ and ${\cal A}$
commute
\begin{equation}
 \left[ {\cal R} ,  {\cal A} \right] = 0
\end{equation}
by virtue of the second heavenly system (\ref{uq}). Moreover, $
{\cal R}$ and $ {\cal A} $ form a Lax pair for the second heavenly
system.

\section{Second and third Hamiltonian structures}
\label{sec-multihamilton}

   The second Hamiltonian operator $J_1$ is obtained by applying the
recursion operator (\ref{recursion}) to the first Hamiltonian
operator $J_1 = {\cal R} J_0$. We find
\begin{equation}
J_{1} =   \left(            \begin{array}{cc}  D_{x}^{\;-1}
 & - \frac{{\textstyle q_x }}{{\textstyle u_{xx} }} \\[4mm]
  \frac{{\textstyle q_x }}{{\textstyle u_{xx} }} &
 \begin{array}{c}
  -\frac{1}{2} \left(Q D_x  \frac{{\textstyle 1
}}{{\textstyle u_{xx} }} + \frac{{\textstyle 1 }}{{\textstyle
u_{xx} }} D_x Q \right)
\\[2mm] + \frac{1}{2} \left( \frac{{\textstyle q_x }}{{\textstyle u_{xx} }}
D_y  \frac{{\textstyle 1 }}{{\textstyle u_{xx} }} +
\frac{{\textstyle 1 }}{{\textstyle u_{xx} }} D_y \frac{{\textstyle
q_x }}{{\textstyle u_{xx} }}
 + \frac{{\textstyle 1 }}{{\textstyle u_{xx} }}
D_z + D_z \frac{{\textstyle 1 }}{{\textstyle u_{xx} }}  \right)
\end{array}
\end{array}  \right)               \label{j1}
\end{equation}
which is manifestly skew. The proof of the Jacobi identity is
again straight-forward and lengthy.

The Hamiltonian operators (\ref{j0}) and (\ref{j1}) form a Poisson
pencil, that is, every linear combination $aJ_0+bJ_1$ of these two
Hamiltonian operators with constant coefficients $a$ and $b$
satisfies the Jacobi identity. This can be verified using the
functional multi-vectors criterion of Olver \cite{olv}.

The vanishing of the total time derivative of the Hamiltonian
functional $ H_0 = \int_{-\infty}^\infty {\cal H}_0 dx dy dz$ with
the Hamiltonian density
\begin{equation}
{\cal H}_0 = (x+c_1) \,q \, u_{xx}, \label{H0}
\end{equation}
where $c_1$ is an arbitrary constant, follows by the divergence
theorem from the fact that $\partial{\cal H}_0/\partial t$,
calculated along the flow (\ref{uq}), is a three-dimensional total
divergence in $x, y, z$ and therefore $H_0$ is an integral of the
motion along the flow (\ref{uq}), provided the components of the
current tend to zero fast enough as $x, y, z$ approach
$\pm\infty$.

The Hamiltonian function ${\cal H}_0$ satisfies the recursion
relation (\ref{hameqpq})
\begin{equation}
u^i_t  =  J^{ik}_0 \delta_k H_1 = J^{ik}_1\delta_k H_0
 \label{biham}
\end{equation}
which shows that the second heavenly equation (\ref{heaven2}) in
the two-component form (\ref{uq}) is a {\it bi-Hamiltonian
system}.

In constructing the second Hamiltonian operator we have used the
fact \cite{magri} that given one Hamiltonian operator $J_0$ and
the recursion operator ${\cal R}$
\begin{equation}
J_n = {\cal R}^n J_0 \label{n}
\end{equation}
is also a Hamiltonian operator. In the case of (\ref{j1}) we have
$n=1$. Now if we act with the recursion operator (\ref{recursion})
on the second Hamiltonian operator $J_1$, or use (\ref{n}) for
$n=2$ we can generate a new Hamiltonian operator $J_2={\cal
R}J_1=J_1J_0^{-1}J_1$ where we have used the fact that by
construction ${\cal R}=J_1J_0^{-1}$. The explicit expression for
$J_2$ is
\begin{equation}
J_{2} =   \left(            \begin{array}{cc} -
D_{x}^{\;-1}D_yD_{x}^{\;-1}
 & -\left(D_x^{\;-1}D_z - \frac{{\textstyle q_y+u_{xz} }}{{\textstyle u_{xx} }}\right)
 \\[8mm]
 D_x^{\;-1}D_z - \frac{{\textstyle q_y+u_{xz} }}{{\textstyle u_{xx} }} &
 \begin{array}{c}
  q_x(q_y+u_{xz})D_x\frac{{\textstyle 1}}{{\textstyle u_{xx}^2 }} +  \frac{{\textstyle 1
}}{{\textstyle u_{xx}^2 }}D_x q_x (q_y + u_{xz})
\\[4mm] - \frac{1}{2} \left(q_x^2D_y \frac{{\textstyle 1 }}{{\textstyle u_{xx}^2 }}
+ \frac{{\textstyle 1 }}{{\textstyle u_{xx}^2 }}D_yq_x^2 \right)
\\[4mm]
-\left( q_xD_z\frac{{\textstyle 1 }}{{\textstyle u_{xx} }} +
\frac{{\textstyle 1 }}{{\textstyle u_{xx} }}D_zq_x  \right)
\end{array}
\end{array}  \right)               \label{j2}
\end{equation}
which is again manifestly skew and we have checked the Jacobi
identity for $J_2$ and its compatibility with $J_1$ and $J_0$.

It can be verified that the Hamiltonian functional $ H_{-1}$ with
the density
\begin{equation}
{\cal H}_{-1} = u - (y+c_2)\, q \, u_{xx}, \label{H-1}
\end{equation}
with $c_2$ another arbitrary constant, is an integral of the flow
(\ref{uq}). Its importance follows from the fact that it also
satisfies the Magri's relation (\ref{hameqpq}) with $J_1$ replaced
by the third Hamiltonian operator $J_2$, so that we obtain a {\it
tri-Hamiltonian representation} of the second heavenly equation
(\ref{heaven2}) in the two-component form (\ref{uq})
\begin{equation}
u^i_t =  J^{ik}_0 \delta_k  H_1
 = J^{ik}_1 \delta_k  H_0
 = J^{ik}_2 \delta_k  H_{-1}.
\label{triham}
\end{equation}
This construction can be continued by the repeated application of
the recursion operator (\ref{recursion}) to Hamiltonian operators.
Thus we have the {\it Magri, or multi-Hamiltonian representation}
of second heavenly system (\ref{uq}).

\section{Symmetries and integrals of motion}
\label{sec-symmetry}

Hamiltonian operators provide a natural link between commuting
symmetries in evolutionary form \cite{olv} and conserved
quantities, integrals of motion, in involution with respect to
Poisson brackets. Our original two-component system (\ref{uq}) is
also a member of the infinite hierarchy of commuting symmetries
and has the form (\ref{triham}) where ${\cal H}_1$, ${\cal H}_0$
and ${\cal H}_{-1}$, defined by (\ref{h1}), (\ref{H0}) and
(\ref{H-1}) respectively, are integrals of motion, while the whole
right-hand side of (\ref{triham}) is the symmetry characteristic
of time translations. If we replace $J_0$ by $J_1$ and $J_1$ by
$J_2$ on the right-hand side of (\ref{biham}) and the time $t$ by
the group parameter $\tau$ we obtain
\begin{equation}
u^i_\tau  = J^{ik}_1 \delta_k H_1 = J^{ik}_2 \delta_k  H_0 = u^i_z
, \label{symz}
\end{equation}
that is, the symmetry of $z$-translations generated either by the
integral $ H_1$ relative to the second Poisson structure $J_1$, or
by the integral $ H_0$ relative to the third Poisson structure
$J_2$. We also note that
\begin{equation}
J_0 \left(
\begin{array}{c}
\delta_u H_0 \\ \delta_q H_0
\end{array}
\right) = J_1 \left(
\begin{array}{c}
\delta_u H_{-1} \\ \delta_q H_{-1}
\end{array}
\right) = \left(
\begin{array}{c}
x+c_3 \\ 0
\end{array}
\right) \label{J0H0}
\end{equation}
where $c_3$ is another arbitrary constant and furthermore
\begin{equation}
J_0 \left(
\begin{array}{c}
\delta_u H_{-1} \\ \delta_q H_{-1}
\end{array}
\right) = - \left(
\begin{array}{c}
y+c_2 \\ 0
\end{array}
\right) \label{J0H-1}
\end{equation}
where $c_2$ is an arbitrary constant coming from $H_{-1}$.

Now we consider the functional $ H^1$ determined in a similar way
to $H_1$ by the density
\begin{equation}
{\cal H}^1 = \frac{1}{2} (u_xu_y - u_x^2q_x). \label{H1}
\end{equation}
and check that the total time derivative of ${\cal H}^1$ along the
flow (\ref{uq}) is a total divergence and hence the functional $
H^1$ is an integral so that it can serve as a Hamiltonian for some
flow commuting with (\ref{uq}). Acting on the column of its
variational derivatives by $J_0$ and $J_1$ we obtain two such
flows
\begin{eqnarray}
u^i_\tau & = & J^{ik}_0 \delta_k H^1 = u^i_x
 \nonumber\\
u^i_\tau & = & J^{ik}_1 \delta_k H^1 = - u^i_y \nonumber
\end{eqnarray}
with symmetry characteristics of $x$ and $y$-translations
respectively.

Now let us present the results of a systematic search for point
symmetries and the corresponding integrals of motion for the
second heavenly system. The complete Lie algebra of point
symmetries of the original one-component heavenly equation
(\ref{heaven2}) was given in \cite{mns} but in the two-component
representation the results look a little different. The basis
generators of one-parameter subgroups of the complete Lie group of
point symmetries for the second heavenly system (\ref{uq}) have
the form
\begin{eqnarray}
& \!\!\!\!\! & X_I = -2z\partial_t + tx \partial_u + x\partial_q ,
\quad X_{II} = t\partial_t + z\partial_z + u\partial_u , \quad W_f
= f(y,z)\partial_u \nonumber
\\ &\!\!\!\!\!\! & X_{III} =
t\partial_t + x\partial_x + 3u\partial_u + 2q\partial_q ,\quad Z_b
= b(y)\partial_z - b^\prime(y)x\partial_t - b^{\prime\prime}(y)
\frac{x^3}{6}\,\partial_u \nonumber
\\ &\!\!\!\!\!\! & Y_a = a\partial_y + a^\prime(x
\partial_x - t\partial_t - z\partial_z + q\partial_q) +
a^{\prime\prime}\!\left(xz\partial_t - \frac{tx^2}{2}\,\partial_u
- \frac{x^2}{2}\,\partial_q\right) \nonumber
\\ &\!\!\!\!\!\! &\mbox{} + a^{\prime\prime\prime}\,\frac{x^3z}{6}\,\partial_u
,\quad V_d = d_z(y,z)(t\partial_u + \partial_q) - d_y(y,z)x
\partial_u
\label{generat}
\\ &\!\!\!\!\!\! & U_c = c_y\partial_t + c_z \partial_x - c_{yz}x(t\partial_u + \partial_q)
+ c_{yy}\frac{x^2}{2}\,\partial_u + c_{zz}\left(\frac{t^2}{2}\,
\partial_u + t\partial_q\right) \nonumber
\end{eqnarray}
where $a(y), b(y), c(y,z), d(y,z)$ and $f(y,z)$ are arbitrary
functions, primes denote ordinary derivatives of functions of one
variable and we used the shorthand notation $\partial_t =
\partial/\partial t$ and so on. Since some of the generators contain
arbitrary functions, the total symmetry group is an infinite Lie
(pseudo)group. In the table of commutators of the generators
(\ref{generat}) the commutator $[X_i,X_j]$ stands at the
intersection of $i$th row and $j$th column and we have used the
shorthand notation $\hat f=zf_z-f$, $\tilde f = af_y - a'zf_z$ and
$B(y)=\int b(y) dy$. Here $\frac{\partial(c,d)}{\partial(y,z)} =
c_yd_z-c_zd_y$ is the Jacobian.
\begin{table}[ht]
\hspace{-2mm}
\begin{tabular}{|c|c|c|c|c|c|c|c|c|}
\hline    &$X_I$&$X_{II}$  &$X_{III}$& $Y_a$ &$Z_b$ &$U_c$ &$V_d$
&$W_f$
\\ \hline
    $X_I$ & $0$ & $0$& $X_I$ & $0$ &$U_{2B}$  &$-V_{2\hat c+c}$   &$-W_{2zd_z}$ &$0$
\\ \hline
    $X_{II}$ &$0$& $0$ & $0$   & $0$ &$-Z_b$&$U_{\hat c}$& $V_{\hat d}$& $W_{\hat f}$
\\ \hline
    $X_{III}$ &$-X_I$  &$0$  & $ 0$  &$0$  &$0$& $-U_c$& $-2V_d$& $-3W_f$
\\ \hline
    $Y_a$ &$0$& $0$& $0$  & $0$& $Z_{(ab)'}$& $U_{\tilde c}$& $V_{\tilde d}$ & $W_{\tilde f}$
\\ \hline
    $Z_b$ &$-U_{2B}$& $Z_b$& $0$& $-Z_{(ab)'}$& $0$& $U_{bc_z}$& $V_{bd_z}$& $W_{bf_z}$
\\ \hline
$U_c$ &$V_{2\hat c+c}$& $-U_{\hat c}$& $U_c$& $-U_{\tilde c}$& $-U_{bc_z}$& $0$& $W_{\frac{\partial(c,d)}{\partial(y,z)}}$&$0$
\\ \hline
$V_d$ &$W_{2zd_z}$& $-V_{\hat d}$& $2V_d$& $-V_{\tilde d}$& $-V_{bd_z}$& $-W_{\frac{\partial(c,d)}{\partial(y,z)}}$& $0$& $0$
\\ \hline
    $W_f$ &$0$   &$-W_{\hat f}$ & $3W_f$& $-W_{\tilde f}$& $-W_{bf_z}$ &$0$  &$0$& $0$
\\ \hline
\end{tabular}
\caption{Commutators of point symmetries of the second heavenly
system.}
\end{table}

We complete this table with the commutation relations
$[Y_a,Y_g]=Y_{ag'-ga'}$, $[Z_b,Z_h]=0$,
$[U_c,U_s]=V_{\frac{\partial(c,s)}{\partial(y,z)}}$, $[V_d,V_e]=0$
and $[W_f,W_r]=0$ where $g=g(y)$, $h=h(y)$, $s=s(y,z)$, $e=e(y,z)$
and $r=r(y,z)$ are arbitrary functions.

We are interested in the integrals of motion generating all these
point symmetries. The relation between symmetries and integrals is
given by the Hamiltonian form of Noether's theorem
\begin{equation}
\left(
\begin{array}{c}
\hat\eta_u \\ \hat\eta_q
\end{array}
\right) =  J_0 \left(
\begin{array}{c}
\delta_u H \\ \delta_q H
\end{array}
\right)
\label{noether}
\end{equation}
where $ H =\int_{-\infty}^{+\infty} {\cal H} dx dy dz$ is integral
of the motion along the flow (\ref{uq}), with the conserved
density ${\cal H}$, which generates the symmetry with the
two-component characteristic \cite{olv} $\hat\eta_u, \hat\eta_q$.
We choose here Poisson structure determined by our first
Hamiltonian operator $J_0$ since we know its inverse $K$ given by
(\ref{kmu}) which is used in the inverse Noether theorem
\begin{equation}
\left(
\begin{array}{c}
\delta_u H \\ \delta_q H
\end{array}
\right) = K \left(
\begin{array}{c}
\hat\eta_u \\ \hat\eta_q
\end{array}
\right)
 \label{noetherinverse}
\end{equation}
determining the integral $H$ corresponding to the known symmetry
$\hat\eta_u, \hat\eta_q$.

We proceed by using the formula (\ref{noetherinverse}) for
reconstructing conserved densities corresponding to all
variational point symmetries. For the symmetry $X_I$ we use its
characteristic $\hat\eta_I=(xt+2zq,x+2zQ)^T$, with $T$ denoting
transposition, in the formula (\ref{noetherinverse}) to obtain the
conserved density
\begin{equation}
{\cal H}_I = (txq+zq^2)u_{xx} +
\left(\frac{x}{2}\,u_x-zu_z\right)u_x . \label{HI}
\end{equation}
For the symmetries $X_{II}$ and $X_{III}$ corresponding integrals
in (\ref{noetherinverse}) do not exist and hence they are not
variational symmetries \cite{olv}. A characteristic of the
symmetry $Y_a$ has the form
\[\hat\eta_a = \left(
\begin{array}{c}
\frac{1}{6}a'''x^3z-\frac{1}{2}a''tx^2-(a''xz-a't)q-a'xu_x-au_y+a'zu_z
\\ a'q-\frac{1}{2}a''x^2-(a''xz-a't)Q-a'xq_x-aq_y+a'zq_z
\end{array}
\right)
\]
and from (\ref{noetherinverse}) a conserved density corresponding
to it is given by
\begin{eqnarray}
&\!\!\!\! & {\cal H}_a =\left\{\frac{\textstyle q^2}{2}(ta'-x z
a'')+q\left[\frac{1}{6}x^3za'''-\frac{1}{2}tx^2a''+a'(zu_z-xu_x)-au_y\right]\right\}u_{xx}\nonumber
\\ &\!\!\!\! &\mbox{}\phantom{H_a =} +
u\left[\frac{1}{2}tx^2a'''-\frac{1}{6}x^3z a^{IV} \right] +
\frac{1}{2} a'' x u_x\left(zu_z-\frac{1}{2}xu_x \right)
\label{Ha}
\\ &\!\!\!\! & \mbox{}\phantom{H_a =} +
\frac{1}{2}a'(zu_yu_z-tu_xu_z-xu_xu_y) - \frac{1}{2} au_y^2
\nonumber
\end{eqnarray}
where $a(y)$ is an arbitrary smooth function. A characteristic of
the symmetry $Z_b$ is
\[\hat\eta_b = \left(
\begin{array}{c}
-\frac{1}{6}x^3b''+xb'q-bu_z
\\ xb'Q-bq_z
\end{array} \right)\]
and from (\ref{noetherinverse}) we obtain the corresponding
conserved density
\begin{eqnarray}
&\!\!\!\! & {\cal H}_b = \left[\frac{\textstyle x}{2}b'q^2-
\left(\frac{\textstyle x^3}{6}b'' + bu_z\right)q \right]u_{xx} +
\frac{1}{6}x^3b'''u-\frac{1}{2}(b'xu_x+bu_y)u_z \label{Hb}
\end{eqnarray}
where $b(y)$ is an arbitrary smooth function. A characteristic of
the symmetry $U_c$ is
\[\hat\eta_c = \left(
\begin{array}{c}
\frac{1}{2}x^2c_{yy}+\frac{1}{2}t^2c_{zz}-txc_{yz}-c_yq-c_zu_x
\\ tc_{zz}-xc_{yz}-c_yQ-c_zq_x
\end{array}
\right)\] and the corresponding conserved density calculated by
(\ref{noetherinverse}) has the form
\begin{equation}
{\cal H}_c = \left[(\sigma-c_zu_x)q-c_y\frac{\textstyle
q^2}{2}\right]u_{xx} - \sigma_yu + \frac{1}{2}\sigma_t u_x^2 +
\frac{1}{2}(c_yu_z-c_zu_y)u_x \label{Hc}
\end{equation}
where $c(y,z)$ is an arbitrary smooth function and we have used
the notation
\[\sigma(t,x,y,z) = \frac{1}{2}t^2c_{zz}-txc_{yz} +
\frac{1}{2}x^2c_{yy} . \] A characteristic of the symmetry $V_d$
is given by $\hat\eta_d = (td_z-xd_y,d_z)^T$ and the corresponding
conserved density is
\begin{equation}
{\cal H}_d = (td_z-xd_y)qu_{xx} - (td_{yz}-xd_{yy})u + \frac{1}{2}
d_zu_x^2 \label{Hd}
\end{equation}
where $d(y,z)$ is an arbitrary smooth function. Finally, the
symmetry $W_f$ has the characteristic $\hat\eta_f = (f, 0)^T$ and
the corresponding conserved density is
\begin{equation}
{\cal H}_f = fqu_{xx} -f_yu \label{Hf}
\end{equation}
where $f(y,z)$ is an arbitrary smooth function.

By a lengthy calculation one may check that the time derivatives
of all these densities ${\cal H}$ along the flow (\ref{uq}) are
total divergences providing an independent check that the
corresponding functionals $ H $ are indeed integrals of motion
subject to suitable boundary conditions. Note that by replacing
$q$ by $u_t$ we obtain integrals of motion for the original form
(\ref{heaven2}) of the second heavenly equation.

Finally, we list some simple obvious symmetries, which are
particular cases of the symmetries $Y_a, Z_b, U_c$ and $V_d$, and
integrals of motion corresponding to them. Among the latter we
find all the Hamiltonian functions ${\cal H}_1, {\cal H}_0, {\cal
H}_{-1}$ and ${\cal H}^1$ (up to inessential arbitrary constants)
which were used so far. Translational symmetries in each of the
four coordinates and integrals of motion generating them are
\begin{eqnarray}
&\!\!\!\!\! & X_t = \partial_t = U_{y}\qquad {\cal H}^t={\cal
H}_{c=y} = -\frac{1}{2}(q^2u_{xx}+u_xu_z) \equiv - {\cal H}_1
\nonumber
\\ &\!\!\!\!\! & X_x = \partial_x = U_z
\label{translations}
\\ &\!\!\!\!\! & {\cal H}^x = {\cal H}_{c=z} = -\left(qu_xu_{xx}+\frac{1}{2}u_xu_y\right) \iff
{\cal H}^x = -\frac{1}{2}\left(u_xu_y-u_x^2q_x\right) \equiv -
{\cal H}^1 \nonumber
\\ &\!\!\!\!\! & X_y =  \partial_y = Y_1 \qquad {\cal H}^y = {\cal H}_{a=1} =
-\left(qu_yu_{xx}+\frac{1}{2}u_y^2\right)\nonumber
\\ &\!\!\!\!\! & X_z = \partial_z = Z_1 \qquad {\cal H}^z = {\cal H}_{b=1} =
-\left(qu_zu_{xx}+\frac{1}{2}u_yu_z\right). \nonumber
\end{eqnarray}
The dilatational symmetry appears as the particular case of $Y_a$
\begin{eqnarray}
& & X^d = Y_y = x\partial_x + y\partial_y - t\partial_t
-z\partial_z + q\partial_q
\label{dilat}
\\ & & {\cal H}^d = {\cal H}_{a=y} = \frac{\textstyle t}{2}(q^2u_{xx}-u_xu_z) +
(zu_z-xu_x-yu_y)\left(qu_{xx}+\frac{1}{2}u_y\right). \nonumber
\end{eqnarray}
Boosts in $z$- and $x$-directions appear as the special cases of
$Z_b$ and $U_c$
\begin{eqnarray}
&\!\!\!\!\! & X^{B_z} = Z_y = y\partial_z - x\partial_t
\label{boost}
\\ &\!\!\!\!\! & {\cal H}^{B_z}
= {\cal H}_{b=y} = \left(\frac{\textstyle
x}{2}q^2-yqu_z\right)u_{xx} - \frac{1}{2} (xu_x+yu_y)u_z\nonumber
\\ &\!\!\!\!\! & X^{B_x} = U_{yz} - X_I = z\partial_t -
y\partial_x \nonumber
\\ &\!\!\!\!\! & {\cal H}^{B_x} = -\left(\frac{\textstyle z}{2}q^2-yqu_x\right)u_{xx}
+ \frac{1}{2} (yu_y+zu_z)u_x\nonumber
\end{eqnarray}
with the latter symmetry combined with $X_I$. Note that the
symmetries $X_I, X_{II}$ and $X_{III}$ are not particular cases of
$Y_a, Z_b, U_c, V_d$ and $W_f$. Shifts in $u$ and $q$ appear as
special cases of $V_d$ and $W_f$
\begin{eqnarray}
&\!\!\!\!\! & X^{s_t} = t\partial_u + \partial_q = V_z \qquad \;\;
{\cal H}_{s_t} = {\cal H}_{d=z} = tqu_{xx} + \frac{1}{2} u_x^2
\nonumber
\\ &\!\!\!\!\! & X^{s_x} = x\partial_u = V_{d=-y} \qquad
\quad  {\cal H}_{s_x} = {\cal H}_{d=-y} = xqu_{xx}\equiv {\cal
H}_0 \label{shift}
\\ &\!\!\!\!\! & X^{s_y} = -y\partial_u = W_{f=-y}
\qquad {\cal H}_{s_y} = {\cal H}_{f=-y} = -yqu_{xx}+u \equiv {\cal
H}_{-1}. \nonumber
\end{eqnarray}

Note that the second heavenly equation itself has divergence form
\begin{equation}
(u_xq_x-u_y)_t=(u_xq_t+u_z)_x \label{divheav2}
\end{equation}
and therefore $h= u_x q_x - u_y $, that is
\begin{equation}
h = {\cal H}_{f=-1} = -qu_{xx} \label{h}
\end{equation}
is a conserved density and can therefore serve as a Hamiltonian.
To find the Hamiltonian flows, we apply the Hamiltonian operators
$J_0$ and $J_1$ to the vector of variational derivatives of the
Hamiltonian functional $ C =\int_{-\infty}^{+\infty} h \, dx \, dy
\, dz$ corresponding to $h$
\begin{equation}
\left(
\begin{array}{c}
\delta_u C\\ \delta_q C
\end{array}
\right) \equiv -\left(
\begin{array}{c}
q_{xx} \\ u_{xx}
\end{array}
\right) \label{varh}
\end{equation}
with the results
\begin{equation}
J_0 \left(
\begin{array}{c}
\delta_u C \\ \delta_q C
\end{array}
\right) = \left(
\begin{array}{r}
-1 \\ 0
\end{array}
\right) \label{J0h}
\end{equation}
and $$ J_1 \left(
\begin{array}{c}
\delta_u C \\ \delta_q C
\end{array}
\right) = \left(
\begin{array}{c}
0 \\ 0
\end{array}
\right)$$
and hence
\begin{equation}
J^{ik}_n \delta_k  C = 0, \qquad n\ge 1 . \label{J1h}
\end{equation} The last formula shows that $ C$ is the Casimir
functional relative to the Poisson structure operators $J_n$ with
$n\ge 1$, that is, these operators have a nontrivial kernel and
therefore are non-invertible on the whole phase space, as opposed
to $J_0$, which implies non-invertibility of the recursion
operator as well. Hence the phase space equipped with any of the
Poisson structures $J_n,\;n\ge 1$ is a Poisson manifold but not a
single symplectic leaf.

\section{Higher flows}
\label{sec-higher}

We know from the work of Fuchssteiner and Fokas \cite{ff} (see
also \cite{sheftel} and references therein) that if a recursion
operator has the form ${\cal R}=J_1J_0^{-1}$, where $J_0$ and
$J_1$ are compatible Hamiltonian operators, then it is hereditary
(Nijenhuis), i.e. it generates an Abelian symmetry algebra out of
commuting symmetries. Moreover, Hermitian conjugate hereditary
recursion operator acting on the vector of variational derivatives
of an integral yields the vector of variational derivatives of
another integral. Therefore we calculate ${\cal R}^\dagger$ to
find
\begin{equation}
 {\cal R}^\dagger =  \left( \begin{array}{cc} (D_x q_x - D_y) D_x^{\;-1}
   & -  D_x Q - D_z  \\[2mm] u_{xx} D_x^{\;-1}
 &  - q_{x}
 \end{array}
\right)  \label{hermit}
\end{equation}
and act by it on the vector of variational derivatives of $ H_1$
with the result
\begin{equation}
{\cal R}^\dagger \left(
\begin{array}{c}
\delta_u H_1 \\ \delta_q H_1
\end{array}
\right) = \left(
\begin{array}{c}
u_zq_{xx}-q_zu_{xx}+2q_xu_{xz}-u_{yz} \\ u_zu_{xx}
\end{array}
\right) = \left(
\begin{array}{c}
\delta_u H_2 \\ \delta_q H_2
\end{array}
\right) \label{delth2}
\end{equation}
where
\begin{equation}
{\cal H}_2 = q \, u_z u_{xx} - \frac{1}{2}u \, u_{yz} \label{h2}
\end{equation}
is a new integral. This also can be checked straightforwardly by
computing the total time derivative of ${\cal H}_2$ along the flow
(\ref{uq}) and showing that it is a total divergence.

By definition ${\cal R}=J_1J_0^{-1}$ we have ${\cal R}^\dagger =
J_0^{-1}J_1$ and therefore (\ref{delth2}) can be rewritten as
\begin{equation}
J^{ik}_0 \delta_k H_2 = J^{ik}_1 \delta_k H_1 = u^i_z. \label{h21}
\end{equation}
A nontrivial result is obtained by applying $J_1$ to the vector of
variational derivatives of $H_2$ which yields
\begin{equation}
\left(
\begin{array}{c}
u_\tau \\ q_\tau
\end{array}
\right) = J_1 \left(
\begin{array}{c}
\delta_u H_2 \\ \delta_q H_2
\end{array}
\right) \equiv \left(
\begin{array}{c}
D_x^{\;-1}D_z (u_xq_x-u_y)-u_xq_z \\[2mm] Q u_{xz} + u_{zz} - q_xq_z
\end{array}
\right) \label{nonlocsym}
\end{equation}
where the right-hand side is a nonlocal symmetry characteristic.
In this case the local integral $H_2$ generates a nonlocal
symmetry.

We obtain similar results using $H^1$ with the density (\ref{H1})
instead of $H_1$. The action of ${\cal R}^\dagger$ on the vector
of variational derivatives of $H^1$ yields again a vector of
variational derivatives
\begin{equation}
{\cal R}^\dagger \left(
\begin{array}{c}
\delta_u H^1 \\ \delta_q H^1
\end{array}
\right) = -\left(
\begin{array}{c}
u_yq_{xx}-q_yu_{xx}+2q_xu_{xy}-u_{yy} \\ u_yu_{xx}
\end{array}
\right) = -\left(
\begin{array}{c}
\delta_u H^2 \\ \delta_q H^2
\end{array}
\right) \label{deltH2}
\end{equation}
where
\begin{equation}
{\cal H}^2 = q \, u_y u_{xx} + \frac{1}{2}u_{y}^2 \label{H2}
\end{equation}
is a new conserved density. Acting by $J_0$ on the variational
derivatives of $H^2$ will not yield new results since ${\cal
R}^\dagger=J_0^{-1}J_1$ and hence $J^{ik}_0 \delta_k H^2 =
-J^{ik}_1 \delta_k H^1 = u^i_{y} $. Using $J_1$ instead of $J_0$
yields
\begin{equation}
J_1 \left(
\begin{array}{c}
\delta_u H^2 \\ \delta_q H^2
\end{array}
\right) = \left(
\begin{array}{c}
D_x^{\;-1}D_y (u_xq_x-u_y)-u_xq_y \\[2mm]
u_{yz}-q_xq_y + u_{xy}Q
\end{array}
\right) \label{nonlocal}
\end{equation}
which is a nonlocal symmetry characteristic generated by a local
integral $ H^2$.

In a similar way we can construct higher integrals and
corresponding higher flows by applying the conjugate recursion
operator ${\cal R}^\dagger$ to the variational derivatives of all
the integrals constructed in section \ref{sec-symmetry}.

\section{Conclusion}

We have cast the second heavenly equation into the form of a
two-component local nonlinear evolutionary system in order to
discover its Hamiltonian structure. We started by presenting its
first Hamiltonian structure. Then we cast the scalar recursion
operator for the second heavenly equation into matrix form. Thus
we were able to obtain explicitly the second and third Hamiltonian
structures for the second heavenly system. The recursion operator
and the operator determining symmetries form a Lax pair for the
two-component system. By Magri's theorem the multi-Hamiltonian
structure makes the second heavenly system a completely integrable
system in four dimensions. Apart from anti-self-dual Yang-Mills
this is first truly completely integrable system in four
dimensions.

Following Magri, we define the (complete) integrability of
Hamiltonian equations as the existence of an infinite number of
conservation laws and symmetries related to them by the Poisson
structure or, equivalently, the existence of a recursion operator
for symmetries. To justify the name `integrable' we should show
how this property will lead to obtaining solutions, that is to the
integration of the system. In our papers \cite{mnsbig,mns} we
suggested the method of partner symmetries for obtaining solutions
of the complex Monge-Amp\`ere equation and the second heavenly
equation of Plebanski. The existence of partner symmetries was
closely related to the existence of the Lax pair of Mason and
Newman \cite{masnew,mw} but there is no known version of the
inverse scattering method in four dimensions which could utilize
this Lax pair in a customary way. A particular choice of partner
symmetries provided differential constraints such that the
equation under investigation together with constraints possessed
an infinite Lie group of contact symmetries which resulted in a
linearizing transformation and exact solutions. Thus we can
consider the method of partner symmetries as an alternative use of
Lax pairs for linearization of the problem as opposed to inverse
scattering method. Our new method for integrating four-dimensional
heavenly equations is still in its early stages of development and
it is not the subject of our study in this paper. We plan to
return to this matter in the future.

We have presented the Lie algebra of all point symmetries and the
integrals of motion which generate all variational point
symmetries of the second heavenly system. We gave examples of some
nonlocal symmetries (higher flows) generated by local Hamiltonian
functions through these Hamiltonian structures. We are in the
process of applying a similar approach to the complex
Monge-Amp\`ere equation.

\section{Acknowledgements}

One of us, MBS, thanks Peter J. Olver for his valuable remark
about the validity of his criteria for checking the Jacobi
identity and Hamiltonian compatibility for matrix
integral-differential Hamiltonian operators. We thank our referees
for their criticism which helped us to clarify some points raised
in our paper.


\begin{thebibliography}{99}
\bibitem{calabi} Calabi E 1954 {\it Proc. Intern. Congr. Math.
Amsterdam} Vol. {\bf 2} p. 206
\bibitem{pleb} Plebanski J F 1975 {\it J. Math. Phys.} {\bf 16} 2395
\bibitem{magri} Magri F 1978 {\it J. Math. Phys.} {\bf 19}  1156; \\
 Magri F 1980 in Nonlinear Evolution Equations and Dynamical Systems
(M. Boiti, F. Pempinelli, and G.Soliani, Eds.), Lecture Notes in
Phys., {\bf 120}, Springer, New York, p. 233
\bibitem{nrma2} Nutku Y 1996 {\it J. Phys. A: Math. Gen.} {\bf 29} 3257
\bibitem{yn} Nutku Y 2000 {\it Phys. Lett. A} {\bf 268} 293
\bibitem{dm2} Dunajski M and Mason L J 2000 {\it Commun. Math. Phys.} {\bf
213} 641
\bibitem{dm} Dunajski M and Mason L J 2003 {\it J. Math. Phys.}
{\bf 44}  3430; \\ math.DG/0301171
\bibitem{mns} Malykh A A
Nutku Y and  Sheftel M B 2004 {\it J. Phys. A: Math. Gen.} {\bf
37} 7527; math-ph/030503
\bibitem{dirac} Dirac  P A M 1964 {\it Lectures on Quantum
Mechanics}Belfer Graduate School of Science Monographs series 2,
New York
\bibitem{galvao} Galvao C A P 1993 private communication.
\bibitem{mnsbig} Malykh A A  Nutku Y and Sheftel M B 2003 {\it J. Phys. A: Math.
Gen.} {\bf 36} 10023; math-ph/0403020
\bibitem{fokas} Santini P M and Fokas A S 1988 {\it Commun. Math. Phys.} {\bf 115}
375
\bibitem{olv}
Olver P J 1986 Application of Lie groups to differential
equations, Springer, New York
\bibitem{ff}
Fuchssteiner B and Fokas A S 1981 {\it Physica} {\bf 4D} 47
\bibitem{sheftel}
Sheftel M B 1996 `Recursions', in: {\it CRC Handbook of Lie Group
Analysis of Differential Equations}, Vol.~{\bf 3}, "New Trends in
Theoretical Developments and Computational Methods", Ch.~4, pp.
91--137, N.H. Ibragimov ed., CRC Press, Boca Raton
\bibitem{masnew}
Mason L J and Newman E T 1989 {\it Commun. Math. Phys.} {\bf 121}
659--668
\bibitem{mw}
Mason L J and Woodhouse N M J 1996 {\it Integrability,
self-duality, and twistor theory} (Oxford: Clarendon Press)
\end{thebibliography}
\end{document}